\documentclass[sigconf]{acmart}
\usepackage{subcaption}
\usepackage{graphics}

\makeatletter
\renewcommand\@formatdoi[1]{\ignorespaces}
\makeatother
\usepackage{booktabs} % For formal tables

\setcopyright{rightsretained}
\acmISBN{} 
\def\acmDOI#1{\def\@acmDOI{#1}} 
\acmDOI{} 

\acmConference[RecSys'18]{Late-Breaking Results track part of the Twelfth ACM Conference on Recommender Systems}{October 2-7, 2018}{Vancouver, BC, Canada}
\acmYear{2018}
\copyrightyear{2018}

\begin{document}

\title{Towards Large Scale Training Of Autoencoders For Collaborative Filtering}
% \titlenote{Produces the permission block, and copyright information}

\author{Abdallah Moussawi}
\orcid{0000-0002-7790-5575}
\affiliation{
  \institution{Anghami}
  \city{Beirut}
  \country{Lebanon}
}
\email{abdallah.moussawi@anghami.com}

% \renewcommand{\shortauthors}{A. Moussawi}

\begin{abstract}
In this paper, we apply a mini-batch based negative sampling method to efficiently 
train a latent factor autoencoder model on large scale and sparse data for implicit feedback
collaborative filtering. We compare our work against a state-of-the-art baseline 
model on different experimental datasets and show that this method can lead to a good
and fast approximation of the baseline model performance. The source code is available \href{https://github.com/amoussawi/recoder}{here}
\footnote{\href{https://github.com/amoussawi/recoder}{https://github.com/amoussawi/recoder}}.
\end{abstract}

% <concept_id>10002951.10003227.10003351.10003269</concept_id>
% <concept_desc>Information systems~Collaborative filtering</concept_desc>

% \ccsdesc[500]{Information systems~Collaborative filtering}


\maketitle

\makeatletter
\newcommand*\bigcdot{\mathpalette\bigcdot@{.5}}
\newcommand*\bigcdot@[2]{\mathbin{\vcenter{\hbox{\scalebox{#2}{$\m@th#1\bullet$}}}}}
\makeatother

\newcommand{\abs}[1]{\left\vert{#1}\right\vert}
\newcommand{\ceil}[1]{\left\lceil{#1}\right\rceil}
\newcommand{\bv}[1]{\textbf{#1}}

\section{Introduction}

% Recommendation systems are an essential component in building 
% personalized web applications. Given a set of items and users, 
% such systems try to recommend a set of items the user hasn't interacted with 
% and that matches his preferences. One successful approach to building 
% recommendation systems is collaborative filtering, which basically generate 
% recommendations based on the patterns in the user-item interactions.

Linear latent factor models \cite{wmf08} are the most popular
collaborative filtering methods in the industry due to their simplicity 
and efficiency. Recent advances have shown that making these models learn
non-linear representations by generalizing them within Autoencoder framework
can achieve better performance \cite{liang18}. 
% These models are optimized to generate top-K recommendations by reconstructing the user-item 
% interactions matrix.
However, one problem with training these models is that it involves 
the reconstruction of highly sparse data with large vector size.
% and it has been shown that models 
% trained with logistic likelihood outperformed those with weighted gaussian likelihood, 
% also models trained with multinomial likelihood has outperfomed models trained with
% either logistic likelihood or gaussian likelihood.

% One good empirical characteristic of training a model with gaussian likelihood
% is that it converges to its best performance in fewer iterations than models
% trained with the other likelihoods \cite{logMF14}, which makes it attractive for industrial
% applications.
In this work, we present a negative sampling method to efficiently train a latent 
factor autoencoder model. This method is based on the simple idea of sampling, for each user,
only the negative items the other users in the mini-batch have interacted with.
% We provide comparative results against a baseline that doesn't use negative
% sampling, and show our model can approximate very well the baseline model performance. 
This is not the first paper to use such method, Hidasi et. al \cite{hidasi15} used a similar 
one but applied for generating recommendations based on short session data with recurrent
neural networks.
% while here applied for generating recommendations based on user full history
% of interactions with latent factor models.

\section{Method}
The user-item interactions matrix is represented as $\textbf{X} \in \{0, 1\}^{\abs{U}\times\abs{I}}$ 
where $U$ and $I$ are the sets of users and items in the dataset, respectively, and 
$\textbf{X}_{u,i} = 1$ if at least one interaction was observed between 
user $u$ and item $i$, otherwise $\textbf{X}_{u,i} = 0$. Given an item $i$
and a user $u$, $U_i$ represents the set of users who interacted with $i$, and 
$I_u$ represents the set of items $u$ has interacted with.

\subsection{Model}
We learn a model $p(\bv{x}_u|\bv{z}_u, \theta) = h(g_{\theta}(\bv{z}_u))$, where $\bv{x}_u = {\bv{X}_{u,*}}^T$ is
the user $u$ vector of interactions, $\bv{z}_u$ is the user latent factor,
$g_{\theta}$ is a multi-layer perceptron parameterized by $\theta$, and $h$ is a function 
that maps the output of $g_{\theta}$ to probabilities based on the likelihood distribution
used to model $p(\bv{x}_u|\bv{z}_u, \theta)$. 
$\bv{z}_u$ can be a learned model parameter \cite{wmf08}, or can be computed as 
a function of $\bv{x}_u$ such that $\bv{z}_u = f_{\lambda}(\bv{x}_u)$, where $f_{\lambda}$ is 
a multi-layer perceptron parameterized by $\lambda$ \cite{liang18}.
The advantage of the second approach over the first is that the model number of 
parameters scales linearly with the number of items only $O(\abs{I})$, whereas in the first it scales linearly with 
both the number of items and users $O(\abs{U} + \abs{I})$. Gaussian likelihood is commonly used to model $p(\bv{x}_u|\bv{z}_u, \theta)$ 
\cite{wmf08}, and two newly studied likelihoods are logistic 
\cite{liang18} and multinomial likelihoods \cite{liang18}. 
The order of increasing performance of those likelihoods is as follows: gaussian, logistic 
and then multinomial \cite{liang18}.

In this paper, we compute $\bv{z}_u = f_{\lambda}(\bv{x}_u)$, such that $g_{\theta} \circ f_{\lambda}$
forms an autoencoder, and we model $p(\bv{x}_u|\bv{z}_u, \theta)$ as a logistic likelihood since it approximates 
the performance of the multinomial likelihood \cite{liang18} and frees us from having a huge softmax at
the output layer. The negative log-likelihood loss function for our model to be minimized is then:
\[- \sum_i \log p(\bv{x}_u|\bv{z}_u, \theta)_i = - \bv{x}_{u} \bigcdot \log(g_{\theta}(\bv{z}_u)) - (\bv{1} - \bv{x}_{u}) \bigcdot \log(\bv{1} - g_{\theta}(\bv{z}_u))\]
To regularize the model, we apply dropout at the input layer and the model is then optimized
to denoise the corrupted version $\Tilde{\bv{x}}_u$ of $\bv{x}_u$ \cite{denoising}, in addition to that
we apply L2 weight decay on $\theta$ and $\lambda$.

% A commonly used loss function is the weighted squared error loss function (Gaussian likelihood) 
% \cite{pmf07, wmf08}, and two newly studied loss functions are the logistic loss 
% function (logistic likelihood) \cite{logMF14, cdae16, liang18} and the multinomial
% distribution negative log-likelihood (multinomial likelihood) loss \cite{liang18}. 

\subsection{Negative Sampling}
For each user in the training mini-batch $\textbf{M} \in \{0, 1\}^{m\times\abs{I}}$, 
where $m$ is the number of users in the batch (batch size), we sample only the negative items 
that the other users in $\textbf{M}$ have interacted with. In other words, we are training the model,
at each mini-batch, only on the set of items that have been interacted with by the users in 
the mini-batch instead of the whole set of items. Such negative sampling 
procedure approximates the sampling from a distribution that is biased towards popular
items, which is a good property to have compared to an unbiased negative sampling. 
A user is more likely to be having no preference for a popular item that he hasn't 
interacted with. However, there are many likely reasons that could make that
user not interact with an unpopular item, such as exposure, or the freshness of the item.
Theoretically, popular items contribute the most to the variance in $\bv{X}$, 
so they should be given more weight when approximating $\bv{X}$'s reconstruction. 

Given that at the start of each training epoch the users are uniformly shuffled,
and given a user $u$ and an item $i$ such that $\textbf{M}_{u,i} = 0$,
the probability that $i$ will be sampled for $u$ is equal to
\[P(\textbf{M}_{*,i} > \textbf{0}) = min(\frac{\abs{U_i}}{N}, 1)\]
where $N = \ceil{\frac{\abs{U}}{m}}$ is the number of training mini-batches. In order 
to tune the sampling probability, one has to tune the batch size $m$. 
Having low batch size $m$ can make the sampling highly biased towards popular 
items, which can lead to overfitting the reconstruction on those popular items. 
On the other hand, having high batch size will sample almost all items, and 
make the sampling obsolete.

The motive behind this sampling method is that it's simple to implement and can 
speed up both forward and backward propagation through the autoencoder.
The idea is, given that $\textbf{M}$ is represented as a sparse matrix in coordinate 
list format, we can sample efficiently the non-zero columns in $\textbf{M}$ 
into a new dense matrix $\textbf{M}_s$ and reconstruct $\textbf{M}_s$.
It can be proven that training the model with this negative sampling procedure
reduces the time complexity of one training iteration over all users 
from $O(\abs{U}\times\abs{I})$ to $O(m\times\sum_{u \in U} \abs{I_u})$, 
so the time complexity of training the model scales linearly with the 
number of interactions in the dataset.

For large datasets with large number of items, we need a large number
of negative samples, hence a large batch size, which makes the batch, 
not fit in memory and expensive to train on. In that case, we 
can simply generate the sparse batch with a large batch size and
then slice it into smaller batches, and train on the small batches.

\section{Experiments}
In our experiments, we follow a similar experimental setup to Liang et. al \cite{liang18}, 
and use their model as a baseline. We experiment with two datasets varying from small-scale 
to large-scale: \textbf{ML-20M} \cite{ml20m}, a movies rating dataset, and \textbf{MSD} \cite{msd}, 
a songs listening count dataset. 
% We filter from the datasets the items that have been interacted 
% with by few users and also the users who have few interactions, and based on that filtering, 
% three variations of the datasets are created as shown in Table \ref{tab:datasets}:
We create three variations of these datasets based on the below filtering. Datasets statistics 
are shown in Table \ref{tab:datasets}:
\begin{itemize}
    \item \textbf{ML-20M}: All ratings above 4 are taken as positive feedback, 
    otherwise they are taken as negative feedback. Only the users who have rated 
    at least 5 movies are kept.
    \item \textbf{MSD}: The users who have listened to less than 20 songs, and 
    the songs that have been listened to by less than 200 users are filtered out.
    \item \textbf{MSD-Large}: The users who have listened to less than 20 songs, and 
    the songs that have been listened to by less than 50 users are filtered out.
\end{itemize}

% \subsection{Datasets}
% Different datasets are used varying from small-scale to large-scale. Table \ref{tab:datasets}
% shows the datasets statistics:
% \begin{itemize}
%     \item \textbf{ML-20M}: A movies rating dataset. All ratings above 4 are 
%     taken as positive feedback, otherwise they are taken as negative feedback. 
%     Only the users who have rated at least 5 movies are kept.
%     \item \textbf{MSD}: A songs listening count dataset. The users who have listened 
%     to less than 20 songs and the songs that have been listened to by less than 
%     200 users are filtered.
%     \item \textbf{MSD-Large}: A variation on MSD dataset above where less songs are filtered
%     out such that the songs that have been listened to by less than 50 users are filtered.
% \end{itemize}
% To evaluate the model we use \textbf{Recall@K} and Normalized discounted cumulative
% gain \textbf{NDCG@K} metrics. 

\begin{table}
  \caption{Datasets statistics}
  \label{tab:datasets}
  \begin{tabular}{cccc}
    \toprule
    & ML-20M & MSD & MSD-Large\\
    \midrule
    \# of users & 136,677 & 571,355 & 629,112\\
    \# of items & 20,108 & 41,140 & 98,485\\
    \# of interactions & 10.0M & 33.6M & 39.7M\\
    sparsity \% & 0.36\% & 0.14\% & 0.064\%\\
    \# of val/test users & 10K & 50K & 50K\\
  \bottomrule
\end{tabular}
\end{table}

\subsection{Setup}
The datasets are split by user into train/validation/test sets. The number of
val/test users are shown in Table \ref{tab:datasets}. For each val/test user,
80\% of his interactions are used to predict the other 20\%, then the model is 
evaluated based on those predictions.

On all datasets, we use the same model architecture, we use an autoencoder with 
a single hidden layer of dimension 200. We set the dropout probability at the 
input layer to 0.5. The weight decay is set to $2e^{-5}$. We optimize the model using
Adam \cite{adam} in batches of 500. For our model, a batch size of 500 was chosen by 
cross-validation since we noticed that for low batch sizes the model was overfitting 
and the validation NDCG@50 metric starts decreasing after few epochs. We train on
ML-20M for 100 epochs, while we train on both MSD and MSD-Large for 80 epochs.

\subsection{Results}
To evaluate the recommendation performance of the model, we use the \textbf{Recall@K} and the 
\textbf{NDCG@K} metrics. In Table \ref{tab:results} 
we compare the recommendation performance of the baseline model versus our model. In Table
\ref{tab:results2} we show the time performance of training both models on 
a CPU and a GPU. The CPU is an 8 cores Intel\textsuperscript{\textregistered} Xeon\textsuperscript{\textregistered} 
E5-2686 v4 and the GPU is a Nvidia\textsuperscript{\textregistered} 
Tesla K80. %The reported time is the time in seconds of training on 10 mini-batches. 
We also show the mean size of the downsampled input vector $\bv{M}_s$ for a batch size of 500, 
the standard deviation of input vector size was less than 800 on all datasets. The size of 
the input vector to the baseline model is always equal to the number of items in the 
dataset irrelevant of the batch size.

It can be seen that our model approximates very well the performance of the 
baseline model with less than 3.72\% decrease in recommendation performance 
while having more than 2.3x speed-up on CPU and more than 2.0x speed-up on GPU.

In future work, we plan to have a thorough comparison of the performance 
of our work versus other fast implementations of the matrix 
factorization (WMF \cite{wmf08}). 
% Preliminary comparison between WMF and our model shows that WMF converges much faster, 
% but, empirically, this is only due to WMF being trained
% with a mean squared error loss, which when used to train our model instead of logistic loss, 
% the model converges much faster than the one trained with logistic loss and 
% has comparable performance with WMF.

\begin{table}
    \caption{Comparison between the recommendation performance of the baseline model and our model.}
    % \centering
    \label{tab:results}
    \resizebox{\columnwidth}{!}{
        \begin{tabular}{ccccccc}
            \toprule
            & \multicolumn{2}{c}{ML-20M} & \multicolumn{2}{c}{MSD} & \multicolumn{2}{c}{MSD-Large}\\
            \midrule
            & Baseline & Ours & Baseline & Ours & Baseline & Ours\\
            \hline
            Recall@20 & 0.3890 & 0.3884 & 0.2677 & 0.2633 & 0.2552 & 0.2453\\
            Recall@50 & 0.5226 & 0.5215 & 0.3540 & 0.3485 & 0.3345 & 0.3233\\
            NDCG@100 & 0.4212 & 0.4193 & 0.3199 & 0.3137 & 0.3059 & 0.2945\\
            \bottomrule
        \end{tabular}
    }
\end{table}

\begin{table}
    \caption{Comparison between the training time performance of the baseline model and our model. 
    The time performance metric is number of batches processed per second.}
    % \centering
    \label{tab:results2}
    \resizebox{\columnwidth}{!}{
        % \begin{tabular}{ccccccc}
        %     \toprule
        %     & \multicolumn{2}{c}{ML-20M} & \multicolumn{2}{c}{MSD} & \multicolumn{2}{c}{MSD-Large}\\
        %     \midrule
        %     & Baseline & Ours & Baseline & Ours & Baseline & Ours\\
        %     \hline
        %     % STD: 367, 490, 765
        %     Input size & 20108 & 5085 & 41140 & 15430 & 98485 & 18740\\
        %     CPU time & 13.07 & 3.28 & 19.64 & 7.73 & 47.3 & 13.81\\
        %     CPU Speed-up & \multicolumn{2}{c}{3.98x} & \multicolumn{2}{c}{2.54x} & \multicolumn{2}{c}{3.42x}\\ 
        %     GPU time & 0.62 & 0.31 & 1.26 & 0.58 & 3.01 & 0.86\\
        %     GPU Speed-up & \multicolumn{2}{c}{2.00x} & \multicolumn{2}{c}{2.17x} & \multicolumn{2}{c}{3.48x}\\ 
        %     \bottomrule
        % \end{tabular}
        \begin{tabular}{ccccccc}
            \toprule
            & \multicolumn{2}{c}{ML-20M} & \multicolumn{2}{c}{MSD} & \multicolumn{2}{c}{MSD-Large}\\
            \midrule
            & Baseline & Ours & Baseline & Ours & Baseline & Ours\\
            \hline
            % STD: 367, 490, 765
            Input size & 20108 & 5085 & 41140 & 15430 & 98485 & 18740\\
            \hline
            CPU rate & 2.1 & 5.6 & 1.1 & 2.5 & 0.5 & 1.6\\
            CPU Speed-up & \multicolumn{2}{c}{\textbf{2.6x}} & \multicolumn{2}{c}{\textbf{2.3x}} & \multicolumn{2}{c}{\textbf{3.2x}}\\
            \hline
            GPU rate & 20.6 & 43.5 & 10.4 & 21.0 & 4.3 & 14.3\\
            GPU Speed-up & \multicolumn{2}{c}{\textbf{2.1x}} & \multicolumn{2}{c}{\textbf{2.0x}} & \multicolumn{2}{c}{\textbf{3.32x}}\\ 
            \bottomrule
        \end{tabular}
    }
\end{table}

\bibliographystyle{ACM-Reference-Format}
\bibliography{bibliography}

\end{document}